\documentclass[footinbib,aps,prl,twocolumn,showpacs,superscriptaddress]{revtex4-1}

\usepackage[margin=1in]{geometry}
\usepackage{amsmath}\usepackage{amsmath} 
\usepackage{amssymb}  
\usepackage{bm}\usepackage{graphicx}
\usepackage[super]{nth}
  \usepackage{hyperref}
  \usepackage[dvipsnames]{xcolor}
 \graphicspath{{images/}}

\newcommand{\pd}{\partial}
\renewcommand{\a}{\alpha}

\renewcommand{\b}{\beta}
\newcommand{\e}{\epsilon}
\renewcommand{\l}{\lambda}
\renewcommand{\r}{\mathbf{r}}
\def\n{\hat{\mathbf{n}}} 
\newcommand{\np}{\hat{\mathbf{n}}_\perp}
\newcommand{\npl}[1]{\hat{\mathbf{n}}_{#1 \perp}}

\newcommand{\bL}{{\boldsymbol\Lambda}}
\newcommand{\be}{\boldsymbol\epsilon}
\newcommand{\bp}{{\mathbf{p}}}
\newcommand{\bq}{{\mathbf{q}}}
\newcommand{\bK}{{\mathbf{K}}}
\newcommand{\bM}{{\mathbf{M}}}
\newcommand{\bd}{{\mathbf{d}}}
\def\beq{\begin{equation}}
\def\eeq{\end{equation}}

\begin{document}

\title{Multi-valued inverse design: multiple surface geometries from one flat sheet}
\date{\today}
\author{Itay Griniasty} \affiliation{Laboratory of Atomic and Solid State Physics, Cornell University, Ithaca, New York 14853-2501, USA}
\author{Cyrus Mostajeran} \affiliation{Department of Engineering, University of Cambridge CB2 1PZ, United Kingdom}
\author{Itai Cohen} \affiliation{Laboratory of Atomic and Solid State Physics, Cornell University, Ithaca, New York 14853-2501, USA} 

\begin{abstract}
Designing flat sheets that can be made to deform into 3D shapes is an area of intense research with applications in micromachines, soft robotics, and medical implants. Thus far, such sheets were designed to adopt a single target shape. Here, we show that through anisotropic deformation applied inhomogenously throughout a sheet, it is possible to design a single sheet that can deform into multiple surface geometries upon different actuations. The key to our approach is development of an analytical method for solving this multi-valued inverse problem. Such sheets open the door to fabricating machines that can perform complex tasks through cyclic transitions between multiple shapes. As a proof of concept we design a simple swimmer capable of moving through a fluid at low Reynolds numbers. 
\end{abstract} 
\maketitle

\begin{figure*}[t]
	\includegraphics[width=\textwidth]{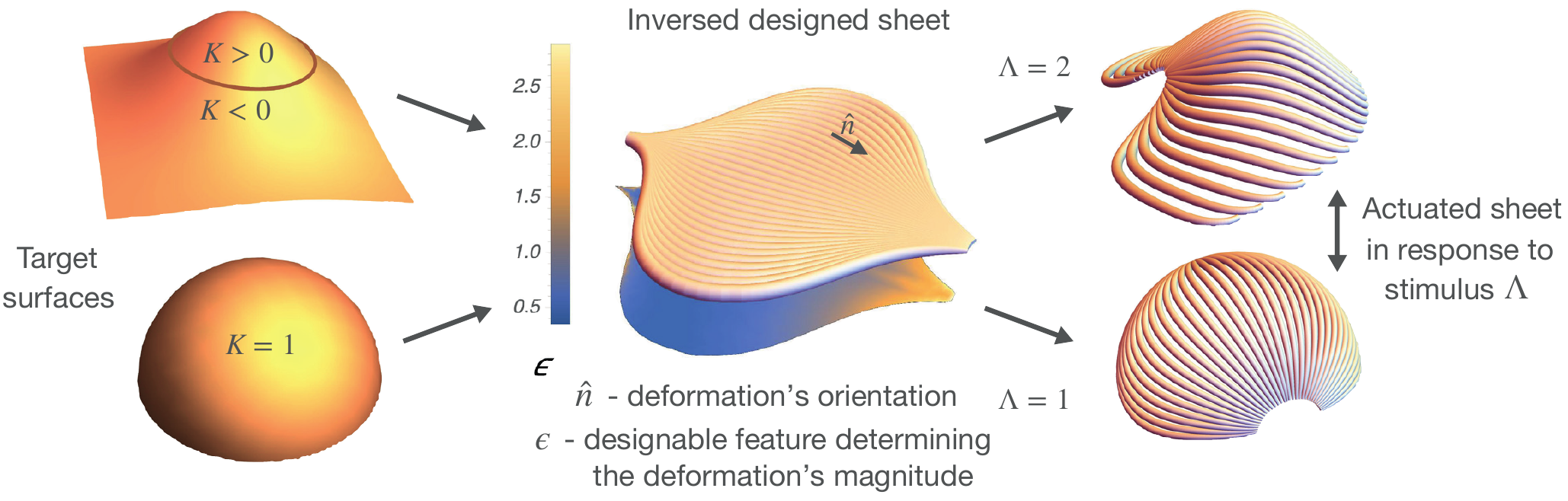}
	\caption{Inverse design of pluripotent sheets. \emph{Left:} Target surfaces with defined Gaussian curvatures \(K\).
	\emph{Center:} Inverse designed sheet. \(\n\) specifies the deformation's orientation. \(\e\) is an experimentally accessible system feature controlling the deformation's magnitude which is designed with \(\n\) such that the sheet deforms into the target surfaces. \emph{Right:} Actuated sheet deforming into multiple target surfaces in response to different values of the stimulus \(\Lambda\).
	}
	\label{fig:shapes}
\end{figure*}

Designing shape shifting sheets is of enormous interest in fields including micromachines \cite{Miskin2020}, soft robotics \cite{RM09,Wallin2018}, and medical implants \cite{Teo2016} where fabrication and production constraints often require an initial flat configuration.
Programming a single shape transformation in such sheets already enables designs for switches, deployable structures \cite{Hartl2007}, and actuators \cite{Guin2018}. Designing sheets that can adopt multiple target geometries would open the door to more sophisticated machines that can cycle through multiple states, perform work on their surroundings, and locomote through viscous fluids \cite{Palagi2018,Levin2020}. 

In origami it is possible to program more than one shape into a single sheet \cite{origamizer, Hayward,qingkun}. However, the shapes are almost always incompatible, which means that the sheet must return to the flat configuration before it is folded into another shape. 
Elastic sheets can fold from one shape to another directly \cite{Balazs20}. Here, we show how to inverse design a sheet so that it can transform into a series of shapes in an arbitrary sequence in response to actuation signals. By transforming from one shape to another directly, without returning to its original flat configuration, such sheets are able to perform complex tasks and do work on their environment.  
The challenge in designing such pluripotent sheets, however, is that one must simultaneously control multiple independent degrees of freedom, such as the deformation magnitude and orientation, to obtain multiple independent shapes (Fig.~\ref{fig:shapes}).

Most shape shifting sheets have been designed to deform into a single target geometry via one of two deformation modalities where only one deformation degree of freedom is varied \cite{MW16}: \((\mathrm{i})\) inhomogeneous isotropic deformations \cite{Korn1914,lichtenstein1916theorie}, where the deformation magnitude varies throughout the sheet \cite{Kim12, Pikul210, KHHS12, KES07}; and \((\mathrm{ii})\) homogeneous anisotropic deformations \cite{gae19}, where the deformation principal axis varies throughout the sheet \cite{FZ55,AEKS11,RM09,AAESK12,ASK14,Mostajeran2015,Gla16,Ware15,War16,Aha18,WM17,War18,SW20,Duffy2020,Feng2020}. Both modalities can be used to alter a sheet's local Gaussian curvature and determine its geometry. 
Importantly, the technology to simultaneously implement both modalities to achieve multiple shapes already exists \cite{Pikul210,BRRS19}. Missing, however, is a mathematical framework to generate designs that implement both degrees of freedom to obtain the desired surfaces.

Naive combinations of the inverse design methods of homogeneous and anisotropic systems \cite{Korn1914,lichtenstein1916theorie,gae19} generally fail at this task.  The naive approach fails because the local Gaussian curvature of an actuated sheet is nonlinearly dependent on both degrees of freedom. The curvature is, however, linear in the highest order derivatives of the deformation degrees of freedom. Therefore, it may be possible to rephrase the inverse design problem as a system of PDEs in the deformation degrees of freedom, where the actuated sheet's curvatures equal those of the target surfaces, and use linearity to simultaneously solve the equations.     

\begin{figure}[t]
	\includegraphics[width=\columnwidth]{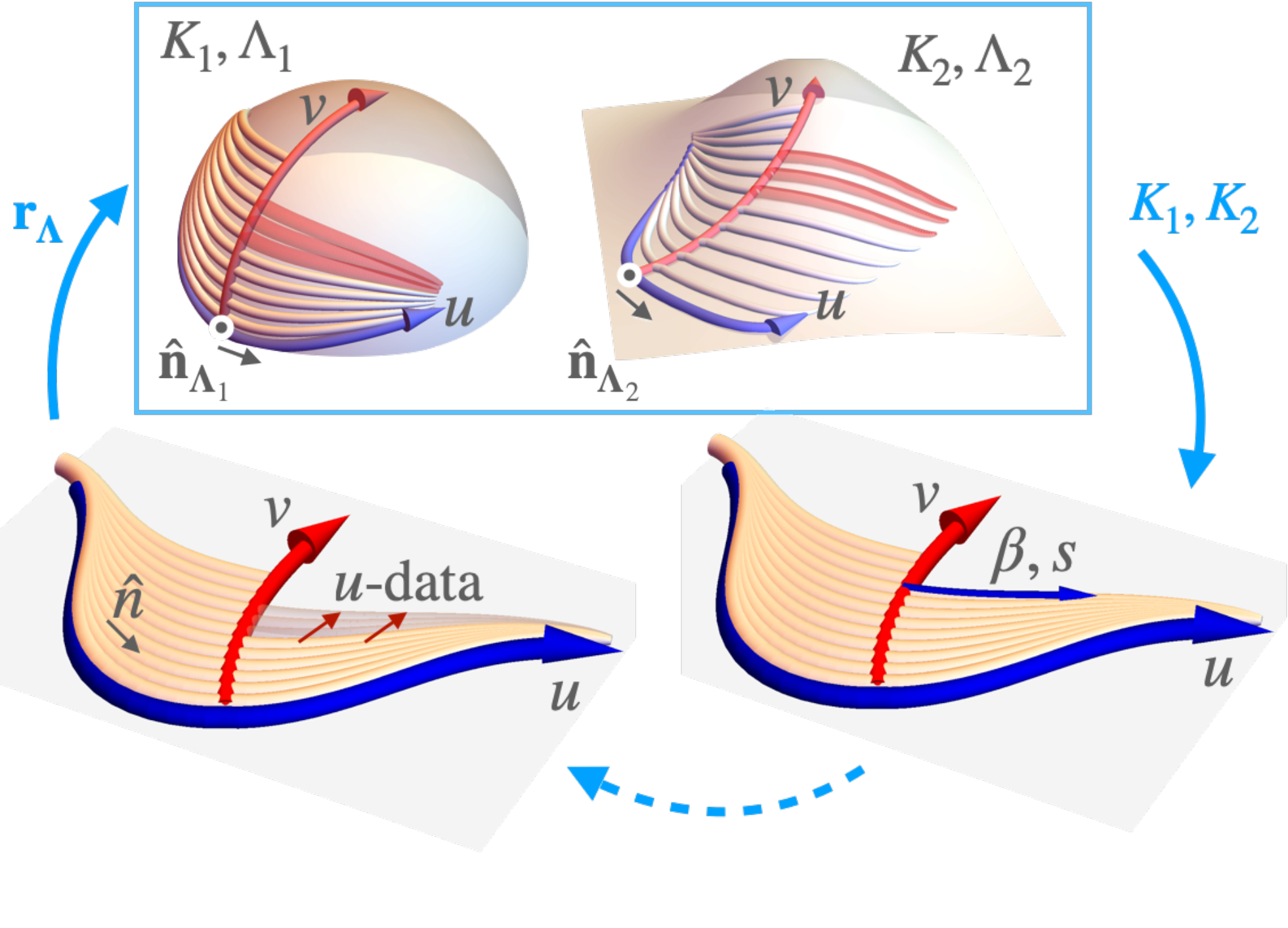}
	\caption{
	Inverse design scheme of a flat uniaxial sheet with a designable elastic response, \(\l_1(\epsilon)\neq\mathrm{const},\) for two target surfaces with Gaussian curvatures \(K_1\) and \(K_2\) in response to stimuli \(\Lambda_1\) and \(\Lambda_2\). 
The \((u,v)\) coordinates given by the integral curves of the deformation's anisotorpy axis \(\n\) or \(\np\), are shared by all surfaces and  and allow the derivation of a  system of equations describing the inverse problem.
A numerical integration of this inverse problem is given by iteration of the following integration steps.
\emph{Bottom left:}  Given complete data on a director integral curve, \(u\text{-data}=(\r_\Lambda,\n_\Lambda,\a,b, \epsilon,q)\) is propagated a step \(dv\) along the director perpendicular, forming a new director integral curve. \emph{Top:} The curvature of the target surfaces along the new director integral curve is obtained. \emph{Bottom right:} Initial data for \(\b\) and \(s\), given on the initial director-perpendicular curve, is integrated along the new director integral curve to complete the data on it. 
}
\label{fig:integration}
\end{figure}

Implementing this strategy requires formulating a common description of the initial and target surfaces in terms of a shared coordinate system. A sheet's deformation is characterized by its principal axis with respect to the initial sheet, denoted by a director \(\n\), and the deformation magnitudes along and across the director \(\l_1\) and \(\l_2\) respectively. 
Generally \(\l_1,\l_2\), and \(\n\) can all depend on external time dependent actuation stimuli \(\bL(t)\) that drive the deformation. Most known anisotropically deforming systems, however, are \emph{uniaxial}: their deformation's principal axis \(\n\) is independent of the actuation \(\bL\) \cite{FZ55,AEKS11,RM09,AAESK12,ASK14,Mostajeran2015,Gla16,Ware15,War16,Aha18,WM17,War18,SW20,Duffy2020,Feng2020,Giudici2021}. In such materials the integral curves of the principal axis \(\n\) and its perpendicular \(\np\) form a `material' coordinate system \((u,v)\) on the sheet throughout the deformation such that 
\begin{equation}
{
\pd_u {\r_{\bL}} =
\a\l_1 {\n_\bL},
\quad
\pd_v {\r_\bL} = 
\b\l_2 \n_{\bL\perp} 
\label{eq:ruv}
}
\end{equation}
where \(\r_\bL\) are the coordinates of the sheet for an actuation \(\bL\),  \(\a\) and \(\b\) are the arc lengths of \(u\) and \(v\) parametric curves on the initial sheet, and  \({\n}_\bL\) are the images of the director \(\n\) on the deformed surfaces (Fig.~\ref{fig:integration}). 
The deformation magnitudes are also functions of designable features \(\boldsymbol{\e}(u,v)\) specified on the undeformed sheet. 
Thus the sheets' geometries throughout the deformation are given by the metrics
\begin{equation}
\label{eq:metric}
ds^2(\mathbf{\Lambda})=\l_1^2(\mathbf{\Lambda},\be)\a^2du^2+\l_2^2(\boldsymbol{\Lambda},\be)\b^2dv^2.
\end{equation}
This shared coordinate system can then be used to define the Gaussian curvatures of multiple actuated surfaces simultaneously. 

An actuated sheet's Gaussian curvature is a function of the deformation degrees of freedom expressed in the metric, Eq.~\eqref{eq:metric}, and its derivatives. For a surface with orthogonal coordinates defined by Eq. \eqref{eq:ruv} the Gaussian curvature is given by \cite{NE17}:
 \begin{equation*}
 {K=\np\cdot\nabla\kappa_{gu}-\n\cdot\nabla\kappa_{gv}-\kappa_{gu}^2-\kappa_{gv}^2}.
 \end{equation*}
where $\kappa_{gu}$ and $\kappa_{gv}$ are geodesic curvatures of \(u\) and \(v\) parametric curves, which are themselves PDEs in the designable director:
\beq
 \label{eq:geodesiccurvature}
 \kappa_{gu}\np=\n\cdot\nabla\n,\quad \quad
 \kappa_{gv}\np=\np\cdot\nabla\n.
 \eeq
Keeping in mind that we would eventually like to design the sheet properties, we express the geodesic curvatures, $\kappa_{gu}$ and $\kappa_{gv}$, as functions of the designable elastic features \(\be\), as well as \(\a\) and \(\b\) which uniquely determine the designable director \(\n\) \cite{NE17} (see supplemental material (SM) for derivation \cite{SM}):
\begin{align}
\label{eq:bspq}
 &\kappa_{gu}=\frac{b}{\l_2}-\frac{\pd\l_1}{\l_1\pd\boldsymbol{\epsilon}}\frac{\mathbf{q}}{\l_2},\quad
\kappa_{gv}=\frac{s}{\l_1}+\frac{\pd\l_2}{\l_2\pd\boldsymbol{\e}}\frac{\mathbf{p}}{\l_1},\\
&b=-\frac{\pd_v\a}{\a\beta},\quad s=\frac{\pd_u\b}{\a\b},\quad\mathbf{p}=\frac{\pd_u\boldsymbol\epsilon}{\a} \text{ and } \mathbf{q}=\frac{\pd_v\boldsymbol\epsilon}{\b}.\notag
\end{align}
Derivatives along and across the director are expressed with respect to \(u\) and \(v\): \(\n\cdot\nabla=\frac{1}{\l_1\a}\pd_u\) and \(\np\cdot\nabla=\frac{1}{\l_2\b}\pd_v\). 
Using Eq.~\eqref{eq:bspq}, we can thus express the Gaussian curvature as a quasi-linear first order equation in \(b,s,\mathbf{p}\) and \(\mathbf{q}\),
\begin{align}
	\label{eq:curvature}
	&K=
	-\left(\frac{s}{\l_1}+\frac{\pd\l_2}{\l_2\pd\boldsymbol{\e}}\frac{\mathbf{p}}{\l_1}\right)^2
	-\left(\frac{b}{\l_2}-\frac{\pd\l_1}{\l_1\pd\boldsymbol{\e}}\frac{\mathbf q}{\l_2}\right)^2+
	\\
	&\frac{1}{\l_2\b}\frac{\pd}{ \pd  v}\left(\frac{b}{\l_2}-\frac{\pd\l_1}{\l_1\pd\boldsymbol{\e}}\frac{\mathbf{q}}{\l_2}\right)
	-\frac{1}{\l_1\a}\frac{\pd}{\pd u}\left(\frac{s}{\l_1}+\frac{\pd\l_2}{\l_2\pd\boldsymbol{\e}}\frac{\mathbf{p}}{\l_1}\right).\notag
	\end{align}
The quantities \(b,s,\bp\) and \(\bq\), and by extension the Gaussian curvature, are thus determined by \(\be\), \(\a\) and \(\b\).

Since \(\boldsymbol\e\), \(\a\) and \(\b\) are independent, the relations in Eq.~\ref{eq:bspq} allow for the simultaneous satisfaction of Eq.~\eqref{eq:curvature} for multiple surface geometries. To obtain solutions, this system of PDEs must be diagonalized, and shown to be integrable. We illustrate this procedure for a flat sheet that deforms into two different target shapes, and show how it naturally extends for an arbitrary number of target shapes.

\emph{Inverse design of two target surfaces:} Consider a uniaxial sheet with a single scalar designable feature \(\e\) affecting the deformation such that, without loss of generality 
\footnote{By definition, \(\e\) is a  designable elastic feature so that either \(\pd_\e\l_1\neq0\) or \(\pd_\e\l_2\neq0\). Without loss of generality we take \(\pd_\e\l_1\neq0\). Otherwise, \(\pd_\e\l_2\neq0\) and by exchanging the director \(\n\) with \(\np\), we rename \(\l_2\) as \(\l_1\).
Further, it is sufficient that only for one value of the actuation \(\Lambda=\Lambda_1\), \(\pd_\e\l_1(\e,\Lambda_1)\neq0\)}, 
\(\pd_\e\lambda_1\neq0\).
The curvatures of the initial sheet and two target surface geometries \(\mathbf{K}=(K_0(\r_{\Lambda_0}),K_1(\r_{\Lambda_1}),K_2(\r_{\Lambda_2}))\) define 3 PDEs in \(\e,\a\) and \(\b\). The equations are linear in \(\pd_vb,\pd_us,\pd_u p\) and \(\pd_v q\).
The variations of \(\e\), \(\pd_u p\) and \(\pd_v q\), however, are not independent, and as shown in the SM \cite{SM} \(\pd_v q\) determines \(\pd_up\) given a Cauchy problem \cite{Hadamard}. 
Recasting Eq.~\eqref{eq:curvature} in terms of the unknown highest order terms
\begin{align}
\label{eq:diagonalization}
\bar{\mathbf{K}}&=\mathbf{K}-\mathbf{M}\cdot \mathbf{d},\\ 
M_i(\Lambda_i)&=
\begin{pmatrix} \frac{1}{\l_2^2}&  -\frac{1}{\l_1^2}& -\frac{\pd\log\l_1}{\l_2^2 \pd\e} \end{pmatrix} ,\quad i\in\{1,2,3\}, \notag \\
\mathbf{d}&=
\begin{pmatrix}  \frac{1}{\b}\pd_v b&\frac{1}{\a}\pd_u s&\frac{1}{\b}\pd_v q\end{pmatrix}, \notag
\end{align}
it is possible to determine \(\pd_v b, \pd_u s\) and \(\pd_v q\)
in terms of \(\bar{\mathbf{K}}\) which are functions of \((\a,\b,\e,p,q,b,s\) and \(\pd_up)\) and the prescribed target curvatures \(\mathbf{K}\): 
\beq
\mathbf{d}=\mathbf{M}^{-1}\cdot ( \mathbf{K}-\bar{\mathbf{K}}).
\label{eq:inverse2}
\eeq
Equations (\ref{eq:ruv},\ref{eq:geodesiccurvature},\ref{eq:bspq}) and (\ref{eq:inverse2}) form a system of PDEs, whose solution is a uniaxial sheet that deforms into the two desired target surfaces upon actuations \(\Lambda_1\) and \(\Lambda_2\) (see Fig.~\ref{fig:shapes}). 
Supplemented by analytical initial conditions, this system is complete and integrable \cite{Evans2010, SM}. Furthermore, if one of the deformation magnitudes is insensitive to \(\e\), which is the case for existing implementations of uniaxial sheets \cite{Pikul210,BRRS19}, 
the system of equations is hyperbolic, and a solution can be integrated from initial conditions for a substantial domain \cite{Recipes07}.

It is illuminating to find solutions of the inverse problem, \(\be(\r)\) and \(\n(\r)\), by integrating a Goursat-like problem \cite{Goursat} as depicted in Fig.~\ref{fig:integration}.
Initial data consists of a position and director on each target surface, accompanied by \(u\) and \(v\) curves on the initial surface, where data propagating across each curve is given on it. That is, data for \((\a,b,\e,q)\) is given on the \(u\)-curve, and data for \((\b,s)\) is given on the \(v\)-curve.
A solution is then found by iteratively propagating the data along \(u\) and \(v\). The variables \(u\text{-data}\equiv\left( \a,\e,b,p ,\pd_vp,q,\{\r_{\bL_i},\n_{\bL_i}\}_{i=0}^2 \right)\) are propagated a step \(dv\), forming a new \(u\)-curve. Next, the curvatures of the target surfaces \((K_1(\r_{\Lambda_1}),K_2(\r_{\Lambda_2}))\) are obtained along the new curve. With these curvatures we obtain the values of \(\pd_u s\) through Eq.~\eqref{eq:inverse2}, which we integrate to obtain \(s\) and \(\b\) along the new \(u\)-curve, completing the data on it. 
The integration steps are iterated until a global solution of the inverse problem, \(\e(\r),\n(\r)\) is found, or until a singularity forms: \(\a=0\), \(\b=0\), or for all applied stimuli \(\l_1=\l_2\).

\emph{Singularities:} The first two singularities where \(\a\) or \(\b\) vanish are defects in the nematic texture discussed in 
\cite{gae19}. The third, \(\l_1=\l_2\) for all applied stimuli, is an isotropic point. At such a point variations of the director no longer affect the deformation, and the sheet cannot be designed to obtain all target curvatures simultaneously. 
The appearance of singularities 
may be delayed by varying the initial conditions, such that  greater coverage of the target surfaces is achieved \cite{gae19}.

\emph{Inverse design of multiple surfaces:}
In general, if there are \(N\) designable features, \(\be\), independently affecting a uniaxial sheet's  deformation, then the sheet may be designed to morph into \(N+1\) independent surfaces. 
Here, the inverse design procedure is nearly identical to that of a sheet morphing into two shapes. 
The key difference is that because there are multiple designable features, we can no longer assume that they all affect the deformations along the director. For example, if \(\e_1\) affects only the deformation along the director, \(\l_1\), while \(\e_2\), affects only the deformation across it, \(\l_2\), then the variations of \(\e_1\) across the director, and \(\e_2\) along it, \(\pd_v q_1\) and \(\pd_u p_2\), are relevant to their inverse design, while \(\pd_u p_1\) and \(\pd_v q_2\) are not.
Equation \eqref{eq:diagonalization} then needs to be modified to account for the relevant highest order terms,  \(\bd\), which now include \(\pd_v b, \pd_us\) and a mix of \(\pd_u p_i\) and \(\pd_v q_j\).  The coefficients matrix \(\bM\) is then appropriately redefined, such that after subtracting \(\bM\cdot \bd\) from the curvature \(\bK\), the remainder \(\bar{\bK}\) is no longer a function of the relevant highest order derivatives.  The accordingly modified
 Eq.\eqref{eq:inverse2},
 together with Eqns.~(\ref{eq:ruv},\ref{eq:geodesiccurvature}) and \eqref{eq:bspq} then compose a complete, integrable system of equations whose solutions are sheets deforming into \(N+1\) target surfaces.
The detailed derivation of the equations and an integration scheme
are given in the SM \cite{SM}.

\begin{figure}[t]
	\includegraphics[width=\columnwidth]{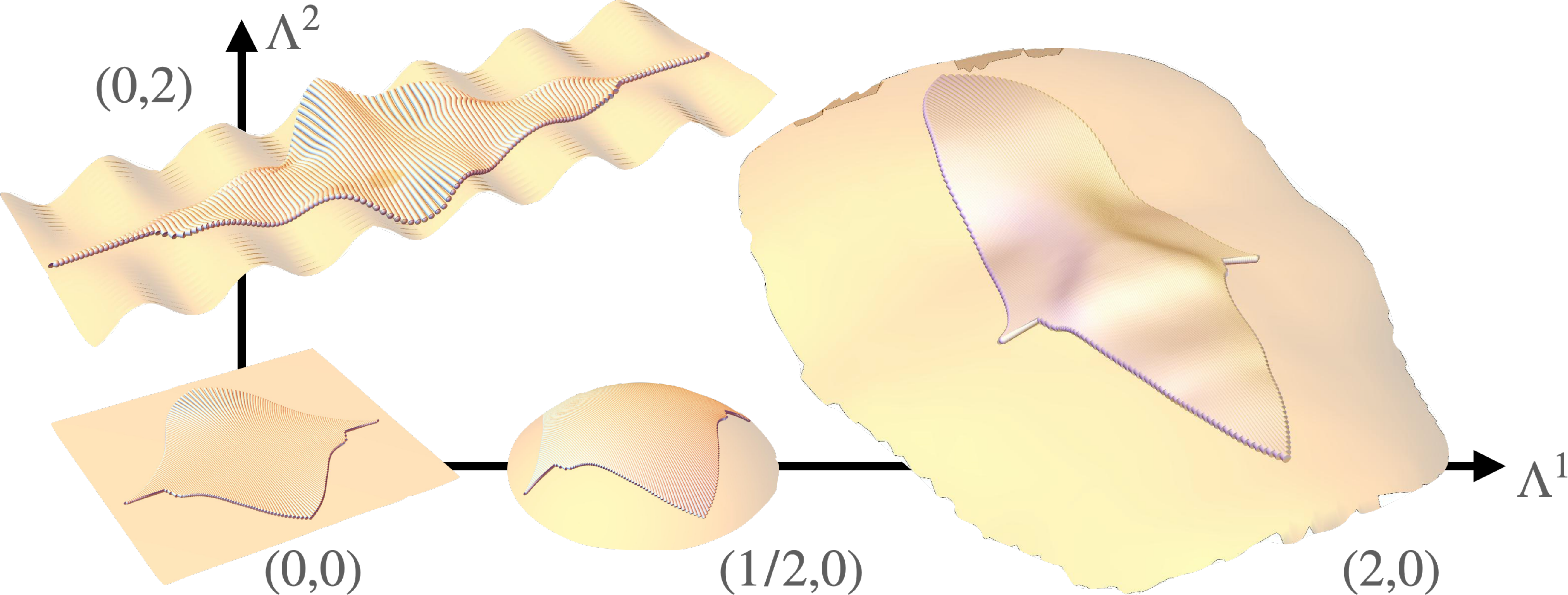}
	\caption{Inverse design of multiple shapes. A flat uniaxial sheet with two designable  system features, \(\l_i=\exp(\e_i \Lambda^i),\,i\in\{1,2\}\), is designed to deform into 3 target shapes in response to two stimuli \(\Lambda^1\) and \(\Lambda^2\). A sphere and face-like mask in response to \(\Lambda^1\) and a wavy sheet in response to \(\Lambda^2\). The maximal strains are below 300\% and are within experimental reach \cite{FNPW01}.
}
	\label{fig:faces}
\end{figure}

\emph{Multiple independent stimuli:} The formulation of the inverse problem and the above integration scheme also hold when the deformation occurs in response to multiple independent stimuli, such as light, pressure or heat, \(\bL=(\Lambda^1,\ldots,\Lambda^k)\).
An example of a solution to such a multi-target inverse problem is presented in Fig.~\ref{fig:faces}. The sheet depicted has two designable features separately affecting its deformation magnitudes in response to independent stimuli: \(\l_1(\Lambda^1,\e_1)\) and \(\l_2(\Lambda^2,\e_2)\).
Such a sheet can morph into highly distinct surfaces. In response to \(\Lambda^1\) the sheet extends along \(\n\) and morphs first into a sphere of constant curvature, and then into a face with a complex curvature profile. In response to \(\Lambda^2\) the sheet morphs across \(\n\) into a surface oscillating along two orthogonal coordinates with two different periods. 
The sheet can then transform into the face, without going through the sphere, by simultaneously changing both stimuli, extending along \(\n\) while contracting across it.
This example illustrates a general feature: the \emph{path} in shape space of a sheet morphing between target geometries in response to multiple independent stimuli can be manipulated in a non-trivial manner.

\begin{figure}[t]
	\includegraphics[width=\columnwidth]{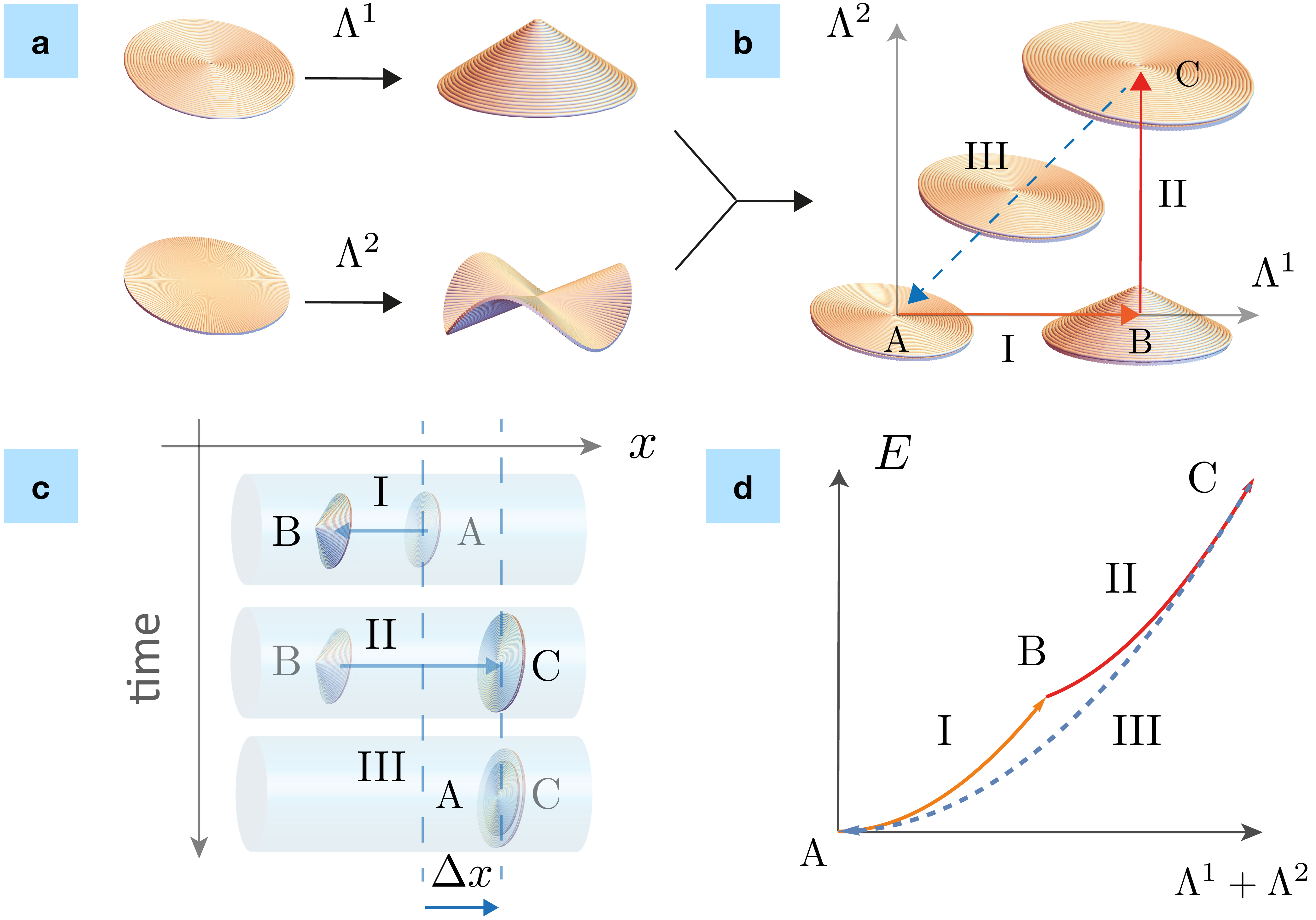}
	\caption{
	(a) A bilayered uniaxial sheet composed of a top layer with an azimuthal director pattern and a bottom layer with the orthogonal radial director pattern. The top and bottom layers deform into cones and anticones upon actuation, respectively \cite{Modes2010a}. 
	(b) A nonreciprocal cycle of shape transformations between multiple flat and conical shapes in the bilayered design through the use of two independently controllable stimuli $\Lambda^1$ and $\Lambda^2$ and the corresponding cycle in parameter space. 
	(c) Locomotion of a swimmer at low Reynolds number with a net displacement $\Delta x$ achieved during each cycle.
	(d) Bounds on the work, $E$ Eq.\eqref{eq:energy1}, that the system can perform on its surroundings at each stage of the cycle.
}
	\label{fig:energy}
\end{figure}

\emph{Locomotion and work:} 
Cycling between multiple shapes is a standard method of doing work and performing complex tasks. It is of particular importance in microscopic machines, such as swimmers, that, due to their size, operate in settings where viscous forces dominate inertial forces resulting in instantaneous flows that have `no memory'. As a consequence only nonreciprocal motions give rise to a net propulsion \cite{Purcell1977}. We provide a simple design for a composite sheet that is capable of locomotion in such environments (Fig.~\ref{fig:energy}).
The composite sheet consists of two homogeneous layers with independently controllable, orthogonal, director patterns, whose actuated Gaussian curvatures are opposite (Fig.~\ref{fig:energy}a).
When the sheets are sequentially actuated and then simultaneously relaxed, they execute a simple non-reciprocal work cycle (Fig.~\ref{fig:energy}b) \cite{Mostajeran2015} that results in an overall translation along the axis of symmetry (Fig.~\ref{fig:energy}c).

This locomotion is powered by the work that the sheet does on its environment. For any actuation, this work is bounded by the frustrated elastic energy \cite{Efrati2009} that would build up if it were constrained to stay in its initial configuration. Because we have defined a common coordinate system we can integrate the energy density along the target surfaces to obtain this bound,
    \begin{equation} \label{eq:energy1}
     \underset{\text{Target Surface}}{E=h \iint \mathcal{E} \left(\lambda_1,\l_2\right)} dA,
    \end{equation}
where \(h\) is the sheet's (unactivated) thickness and the energy density \(\mathcal{E}\) on a target surface is derived in the SM \cite{SM}. Finally, while we have used all the deformation degrees of freedom to obtain the target surfaces, we can still vary the initial conditions to control the sheet's capacity to do work along a prescribed curve on the target surface. Such control may find applications in the design of lifters \cite{White2018} for instance, where a greater concentration of elastic energy at points of contact may be advantageous. Collectively, the ability to use this inverse design approach to design a sheet that can morph into multiple surfaces capable of executing locomotion and even concentrating elastic energy at specific locations is quite remarkable. 

\emph{Discussion.}
By systematically utilizing multiple degrees of freedom to program a single sheet of material so that it can transform into multiple target geometries, we have provided a vital theoretical foundation for the design of printable sheets capable of executing complex behaviors. The inverse design of a specific system using this approach is straightforward. One needs to: i) specify the set of designable features and how they affect the deformation magnitudes in response to stimuli; ii) 
select a compatible number of target surfaces to be obtained at specified actuation values; and iii) choose initial conditions for the integration procedure. Using these inputs the code \cite{code} produces designs for the director \(\n\) and designable features \(\be\). 

Candidate systems for the implementation of this design modality include: (i) Liquid crystal elastomers, where the deformation's orientation and magnitude can be controlled by  varying the nematic director's in-plane and out-of-plane orientation \cite{Ware15,Auguste18}, or the extent of deformation in response to the nematic phase transition \cite{Kuenstler2020}. (ii) 4D printed hydrogels, anisotropically deforming along aligned cellulose fibrils, whose orientation in and out of the plane similarly control the deformation's orientation and magnitude. 
(iii) Micro robotic kirigami metamaterial sheets where the deformation's orientation and magnitude can be controlled by varying the local bending of chemical or electrochemical actuators \cite{Liu21,Miskin2020, Miskin18, Baris20}.
In each of these examples, the deformations are typically applied globally.
As fabrication techniques improve, it may be possible to control the actuation at each point along the surface independently. In this scenario we can use the inverse design framework developed here to obtain the desired target shape by treating the actuation as a designable feature. 
Such designable local actuations would allow a single sheet to update its target curvatures on the fly and morph into almost any desired surface geometry in real time.

\textbf{Acknowledgments }
We thank James Sethna for insightful discussions. 
This work was supported by the Army Research Office (ARO W911NF-18-1-0032), the National Science Foundation (EFMA-1935252) the Cornell Center for Materials Research (DMR-1719875). I.G. also received partial support from the Cornell Laboratory of Atomic and Solid State Physics. C.M. was supported by Fitzwilliam College and a Henslow Research Fellowship from the Cambridge Philosophical Society.


%

\pagebreak
\widetext
\begin{center}
	\textbf{\large Supplemental material - Multi-valued inverse design: multiple surface geometries from one flat sheet}

\end{center}
\setcounter{equation}{0}
\setcounter{figure}{0}
\setcounter{table}{0}
\setcounter{page}{1}
\makeatletter
\renewcommand{\theequation}{S\arabic{equation}}
\renewcommand{\thefigure}{S\arabic{figure}}
\renewcommand{\bibnumfmt}[1]{[S#1]}
\renewcommand{\citenumfont}[1]{S#1}

\section{Integrating the Inverse design of a uniaxial sheet deforming into \(N\ge 2\) surfaces}

Let us collect the evolution equations derived in the body of the text with respect to the material coordinates  \((u,v)\).
The position \(\r_{\bL}\) on the different surfaces evolves with the material coordinates according to
\begin{align}
\pd_u \r_{\bL_i} &=\a\l_1 {\n_{\bL_i}},\quad\pd_v \r_{\bL_i} =\b\l_2 \npl{\bL_i}
\label{eqSM:ruv}
\end{align}
The images of the director on the  different surfaces evolves with the material coordinates according to
\( \kappa_{gu}\np=\n\cdot\nabla\n,\,
 \kappa_{gv}\np=\np\cdot\nabla\n\).
Recalling \(\nabla_{j\bL} n^i_\bL=\pd_j n^i_\bL+\Gamma^i_{\bL jk} n^k_\bL\), with \(\Gamma_{\bL\, jk}^i\) the Levi-Civita connection of the relevant surface, we can write
\begin{align}
 \label{eqSM:dn}
\pd_u\n_{\bL}&= {\a\l_1(\bL)}\left(\kappa_{gu} (\bL)\npl{\bL}-\n_{\bL}\cdot\Gamma_{\bL}\cdot\n_{\bL}\right)\\
\pd_v\n_{\bL}&= {\b\l_2(\bL)}\left(\kappa_{gv} (\bL)\npl{\bL}-\npl{\bL}\cdot\Gamma_{\bL}\cdot\n_{\bL} \right)\notag
 \end{align}
The geodesic curvatures of \(u\) and \(v\) curves are given by
 \begin{align}
\label{eqSM:bspq}
 &\kappa_{gu}=\frac{-\pd_v \log(\a \l_1)}{\b\l_2}=\frac{b}{\l_2}-\frac{\pd\l_1}{\l_1\pd\epsilon}\frac{q}{\l_2},\\
&\kappa_{gv}=\frac{-\pd_u \log(\b \l_2)}{\a\l_1}=\frac{s}{\l_1}+\frac{\pd\l_2}{\l_2\pd\epsilon}\frac{p}{\l_1},\notag\end{align}
where
\begin{align}
&b=-\frac{\pd_v\a}{\a\beta},\quad s=\frac{\pd_u\b}{\a\b},\quad\mathbf{p}=\frac{\pd_u\boldsymbol\epsilon}{\a} \text{ and } \mathbf{q}=\frac{\pd_v\boldsymbol\epsilon}{\b}.\notag
\end{align}
We reinterpret \(b,s,\mathbf{p}\) and \(\mathbf{q}\) as propagations of the designable system properties:
 \begin{align}
\pd_v\a&=-\a\b b,\\
\pd_u\b&=\a\b s,\notag\\
{\pd_u\boldsymbol\epsilon}&={\a}\mathbf{p} \text{ and }\notag\\
{\pd_v\boldsymbol\epsilon}&=\b\mathbf{q}. \notag
\end{align}
Finally, to find propagation equations for \(b,s,\mathbf{p}\) or \(\mathbf{q}\) we cast the Gaussian curvatures as a function of their derivatives:
	\begin{align}
	\label{eqSM:curvature}
	K&=
	\frac{1}{\l_2 ^2 \beta  }\pd_v b
	- \frac{1}{ \l_1 ^2 \alpha }\frac{\pd \log\l_2 }{\pd \epsilon _i} \pd_u p_i
	- \frac{1}{\l_2 ^2 \b}\frac{\pd \log\l_1 }{\pd \epsilon _i} \pd_v q_i
	- \frac{1}{\l_1 ^2\alpha  }\pd_u s \\
	& -\frac{b^2}{\l_2 ^2}-\frac{s^2}{\l_1 ^2}
	-\frac{1}{\l_2\l_1}\frac{\pd}{\pd\epsilon_i}\left(\frac{1}{\l_1}\frac{ \pd\l_2}{\pd\epsilon_j}\right)p_i p_j
	-\frac{1}{\l_2\l_1}\frac{\pd}{\pd\epsilon_i}\left(\frac{1}{\l_2}\frac{ \pd\l_1}{\pd\epsilon_j}\right)q_i q_j 
	+\frac{\pd \log (\l_1^2/\l_2)}{\l_2^2\pd \epsilon_i}b q_i \notag
	+\frac{\pd \log (\l_1/\l_2^2)}{\l_1^2\pd \epsilon_i}s p_i ,	\notag
	\end{align}
where we have assumed Einstein's summation convention.
Extracting the relevant highest order terms
\begin{align}
\mathbf{d}&=
\begin{pmatrix}  \frac{1}{\b}\pd_v b&\frac{1}{\a}\pd_u s&dw_1&\ldots&dw_N\end{pmatrix} ,\notag\\
dw_i&= \begin{cases}
\frac{1}{\a} \pd_u p_i & \text{ if } \exists j\text{ s.t } \partial \l_2(\bL_j)/\partial \epsilon_i \neq 0\\
\frac{1}{\b} \pd_v q_i & \text{ else }
 \end{cases}, \notag
\end{align}
we define the coefficients matrix
\begin{align}
M_i(\bL_i)&=
\begin{pmatrix} \frac{1}{\l_2^2},&  -\frac{1}{\l_1^2},& -\frac{1}{\l_1\l_2}\frac{\pd\l}{\bar{\l} \pd\e_1},&\ldots,&\frac{1}{\l_1\l_2}\frac{\pd\l}{\bar{\l} \pd\e_N} \end{pmatrix} ,
\quad i\in\{1,\ldots,N+2\}
\\
\frac{\pd\l}{\bar{\l}\pd\epsilon_{i}}&= \begin{cases}
\frac{\pd\l_2}{\l_1\pd\epsilon_{i}} & \text{ if } \exists j\text{ s.t } \partial \l_2(\bL_j)/\partial \epsilon_i \neq 0\\
\frac{\pd\l_1}{\l_2\pd\epsilon_{i} }& \text{ else }
 \end{cases} \notag
 \label{eqSM:M}
\end{align}
such that 
\(
\bar{\mathbf{K}}=\mathbf{K}-\mathbf{M}\cdot \mathbf{d}
\)
is independent of \(\bd\), and the inverse design equations of \(b,s\) and \(\bp\) or \(\bq\) are given by
\beq
\bd=\bM^{-1}\cdot(\bK-\bar\bK).
\label{eqSM:inverse}
\eeq

\subsection{Integrability} 
The system of equations includes equations along both coordinates \(u\) and \(v\) for \(\r,\n\) and \(\be\). 
The equations are commensurate, \(\pd_{u}\pd_{v}\r=\pd_{v}\pd_{u}\r\), and  \(\pd_{u}\pd_{v}\n=\pd_{v}\pd_{u}\n\), as has been shown in \cite{gae19}. To show the integrability of the variations of \(\be\) given
in Eq.~\eqref{eqSM:bspq}, consider
\begin{align*}
    \pd_u\pd_v\be&=\pd_u(\b \bq)= \a\b s \bq + \b \pd_u \bq,\\
    \pd_v\pd_u\be&=\pd_v(\a \bp)= -\a\b b \bp + \a \pd_v \bp ,
\end{align*}
from which we derive an equation on the initial condition that is preserved through the propagation along \(u\) or \(v\):
\begin{equation}
\label{eqSM:compatability}
\frac{1}{\a}\pd_u\bq-\frac{1}{\b}\pd_v\bp =b\bp+s\bq.
\end{equation}
We can then reduce the set of equations by looking at the difference in the propagation of  \(\r,\n\) and \(\be\) along \(u\) and along \(v\).
Changing coordinates to \(x_0=u-v\), and \(x_1=u+v\), the system of equations, and analytical initial conditions given along a curve, satisfy the conditions of the Cauchy-Kovalevskaya theorem \cite{Kovalevskaya}. Thus, the system is integrable and a solution may be found within a local environment of the initial curve.

Let us complete this section by noting that finding \(\pd_v \bq\) of \(\pd_u \bp\) (or the inverse) is compatible with a Cauchy problem.
A Cauchy initial condition includes \(\be\) on a non-characteristic curve \(\r(l)\) and its derivative across the curve \(\pd_{l_\perp}\be\), as well as the arc-lengths \(\a\) and \(\b\). The equations \(\pd_l\pd_{l_\perp}\be = a_1(l)\) and  \(\pd_l\pd_{l}\be = a_2(l)\) are two linear equations in \(\pd_u \bp,\,\pd_v \bp, \pd_u \bq\) and \(\pd_v \bq\), which together with Eq.~\eqref{eqSM:compatability} can be used to find an equation for \(\pd_v\bq(\pd_u\bp)\).

\subsection{Integrating the system of PDEs}
\emph{Initial conditions:}
We show here how to integrate a Goursat-like problem, where the initial conditions are the same as those specified in the main text except for the assignment of initial values of \(\be\) and their derivatives. 

The initial conditions consist of: initial positions and directors on both initial and target surfaces,
two perpendicular twice differentiable curves on the initial surface emanating from the initial position, assigned respectively as the initial  \(u\) and \(v\) curves. On the \(u\)-curve we give \(\a(u,0)\), and on the \(v\)-curve \(\b(0,v)\). 
Finally, for designable features denoted \(\e_i^{(v)}\), where \(dw_i=\frac{1}{\a}\pd_u p\), the initial values \(\e_i^{(v)}\) and \(p_i^{(v)}\) are assigned along the initial \(v\)-curve, while for designable features denoted \(\e_i^{(u)}\), where \(dw_i=\frac{1}{\b}\pd_v q\), initial values for \(\e_i^{(u)}\) and \(q_i^{(u)}\) are assigned along the initial \(u\)-curve.

\emph{Initialization:}
The initial data does not explicitly specify the values of all the defined variables appearing in the system of equations that describe the inverse problem.
These are \(\r_\bL,\n_\bL,\a,\b,\be,b,s,\bp,\bq\). To complete the data we need to find \(b,s\) and the \(\bp,\bq\) when not prescribed.
    
For designable features \(\e_i\) given on the initial \(u\)-curve (\(v\)-curve), their variation, and the value of \(\a\) (\(\b\)) along the curve complete the missing data for \(p_i\) (\(q_i\)).
The bend and splay are respectively the geodesic curvatures of the initial \(u\) and \(v\) curves given by Eq.~\eqref{eqSM:bspq}. Thus, there is complete data at \((u,v)=(0,0)\).

\emph{Data propagation:}
It is convenient to propagate the data along diagonals in the \((u,v)\)-plane, as such propagation preserves the \(u\)-\(v\) symmetry of the equations. We implement this approach in the following integration scheme.

The data is divided into sets propagating along \(u\) and \(v\).
\(u\)-data\(=\{\a,b,\bq^{(u)}\}\) propagate along \(v\), \(v\)-data\(=\{\b,s,\bp^{(v)}\}\) propagate along \(u\), \(\r_\bL,\n_\bL\) and \(\e\) propagate along both.
Given complete data along a diagonal (such as the origin), \(u\)-data is propagated a step \(dv\), \(v\)-data is propagated a step \(du\), and \(\r_\bL,\n_\bL\) and \(\e\) are propagated along either. 
The intersections of the diagonal with the \(u\) and \(v\) curves contain the missing \(u\) and \(v\) data.
Finally, we complete the data on the new diagonal by propagating \(\be\) onto the next diagonal to derive \(\bp^{(u)}\) and \(\bq^{(v)}\).

\begin{figure}[t]
    \includegraphics[width=\columnwidth]{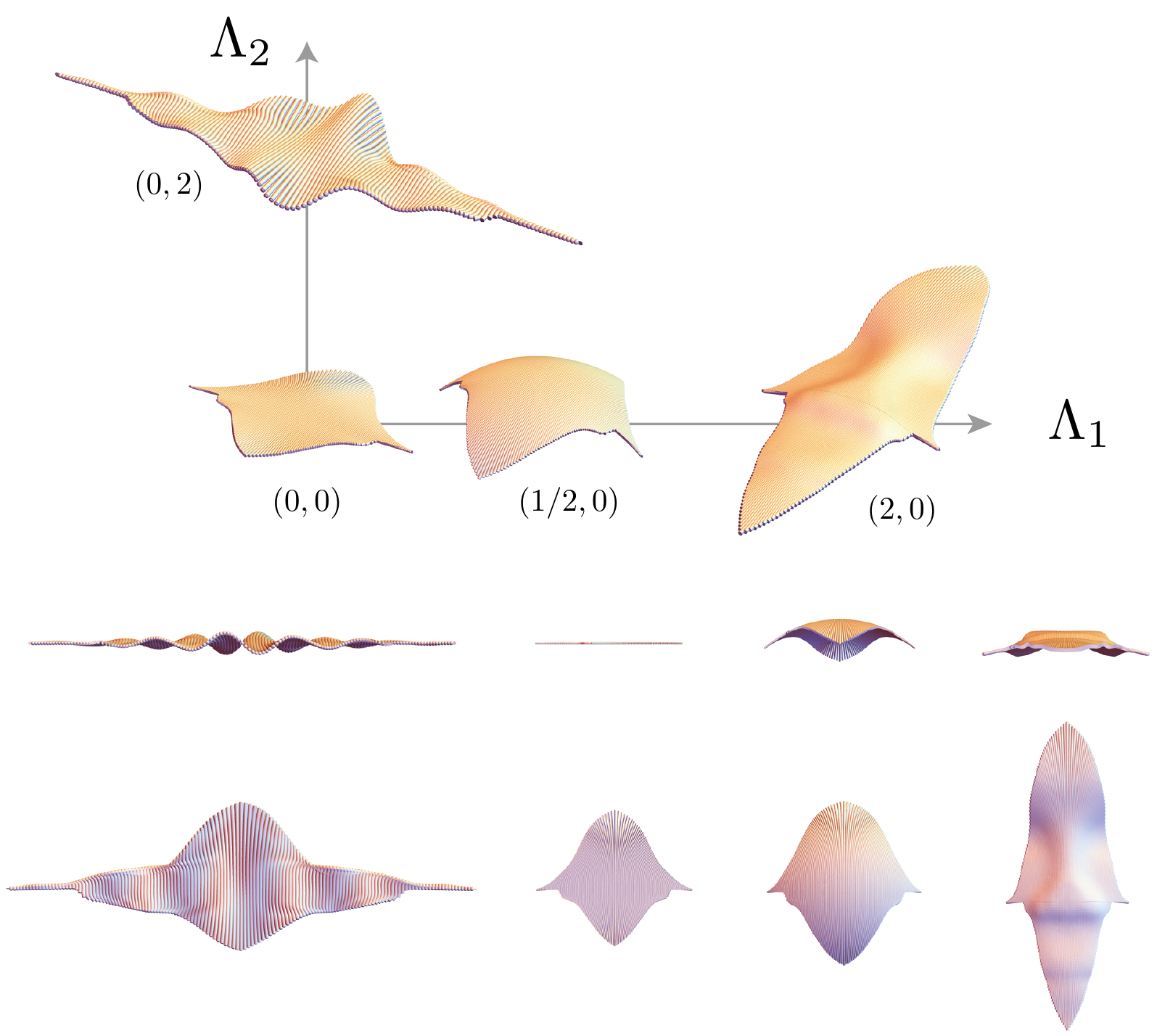}

	\caption{Inverse design of multiple shapes. A flat uniaxial sheet with two designable  system features, \(\l_i=\exp(\e_i \Lambda^i),\,i\in\{1,2\}\), is designed to deform into 3 target shapes in response to two stimuli \(\Lambda^1\) and \(\Lambda^2\). A sphere and face-like mask in response to \(\Lambda^1\) and a wavy sheet in response to \(\Lambda^2\).
	\emph{Top:} Perspective view of the actuated sheet, shown along the stimuli axes. \emph{Middle:} Front view, demonstrating the variations of the geometries along the \(z\) axis. \emph{Bottom:} Top view, demonstrating the mapping of the boundaries, and their underlying diagonal structure in the \(u,v\) coordinates.
}
	\label{fig:shapesSM}
\end{figure}

This integration scheme is implemented to inverse design the sheet presented in Fig.~\ref{fig:shapesSM} which is the same as Figure 3 of the main text. Fig.~\ref{fig:shapesSM} illustrates the diagonal boundaries in \((u,v)\) whose image we see in the top view of the actuated sheet. Singularities in the integration lead to defects which propagate along the diagonals.

\subsection{A word of caution on the choice of actuation sets}
The four-surface inverse problem is set up by attributing a pair of values to the actuation parameters \(\{(\Lambda_1^{(i)},\Lambda_2^{(i)})\}_{i=0}^3\) for each of the surfaces.
Surprisingly, some sets of actuation pairs systematically do not lead to an integrable set of PDEs for 
\(\{\a,\b,\e_1,\e_2\}\).
This is captured by \(\mathrm{det}(\bM)=0\), that is, a linear dependence of the  target curvatures via \(\bar\bK\) on the highest order derivatives \(\bd\).
One such example is \((\l_1,\l_2,K)\in\{(0,0,K_0),(\Omega_1,0,K_1),\,(\Omega_1,\Omega_2,K_2),\,(0,\Omega_2,K_3)\}\). The induced system of equations are 3 independent PDEs of second order in \(\a,\b,\e_1\) and \(\e_2\) given the curvatures \(\{K_0,K_1,K_2\}\) and a first order PDE given all four curvatures, composing a system of PDEs with a constraint whose solution is beyond the scope of this paper. We note that the manifold of actuation pairs with vanishing determinants is of co-dimension \(1\), and so, rare.

\section*{Elastic energy calculations in frustrated uniaxial systems}

The total elastic energy in the reduced 2D model of non-Euclidean elasticity theory \cite{Efrati2009} takes the form
\begin{equation}
E=hE_S + h^3E_B,
\end{equation}
where $h$ is the thickness of the sheet and 
\begin{equation} \label{stretch and bend content}
E_S=\int w_S\sqrt{|\bar{g}|}dx^1dx^2, \quad  \quad \quad E_B=\int w_B\sqrt{|\bar{g}|}dx^1dx^2
\end{equation}
are the stretching and bending contents of the energy determined by
\begin{align} \label{w}
w_S &= \frac{Y}{8(1+\nu_e)}\left(\frac{\nu_e}{1-\nu_e}\bar{g}^{ij}\bar{g}^{kl}+\bar{g}^{ik}\bar{g}^{jl}\right)(g_{ij}-\bar{g}_{ij})(g_{kl}-\bar{g}_{kl}) \\
w_B &= \frac{Y}{24(1+\nu_e)}\left(\frac{\nu_e}{1-\nu_e}\bar{g}^{ij}\bar{g}^{kl}+\bar{g}^{ik}\bar{g}^{jl}\right)b_{ij}b_{kl}.
\end{align}
Here $\bar{g}$ denotes the 2D \emph{reference} metric (i.e. the metric of the target surface) and $g$ the \emph{actual} metric. Integration in Eq.~\eqref{stretch and bend content} is with respect to the reference metric and the indices run over $1$ and $2$ using Einstein notation. $b$ denotes the second fundamental form of the realized surface. $Y$ denotes the Young's modulus and $\nu_e$ the Poisson ratio of the material.

For thin sheets, the bending contribution of the energy is dominated by the stretching contribution as $h^3 \approx 0$. In the thin sheet limit, the actual metric will match the reference metric of the target surface in the unconstrained problem so that the stretching contribution becomes zero and the final configuration is determined by an isometric embedding of the metric that minimizes the bending energy. On the other hand, if the system is constrained (i.e. stretching is blocked somehow), the stretching energy will quickly build up as the actual metric is prevented from matching the reference metric and the bending energy will be negligible in comparison. This stretching energy will then provide an upper bound on the amount of energy that can be extracted from the activation of such a surface to do work, e.g. as lifters \cite{Guin2018}.

Considering only the stretching contribution to the elastic energy of a constrained activated surface, we can express the elastic energy as
\begin{equation} \label{trace SM}
E=\frac{hY}{8(1-\nu_e^2)}\int\left[(1-\nu_e)\operatorname{Tr}\left(\bar{g}^{-1}(g-\bar{g})\right)^2+\nu_e\operatorname{Tr}^2\left(\bar{g}^{-1}(g-\bar{g})\right)\right]\sqrt{|\bar{g}|}dx^1dx^2,
\end{equation}
using the trace operator $\operatorname{Tr}$.
Working in the $(u,v)$-coordinate system outlined in the paper, the pre-actuated metric takes the form $ds^2 = \alpha^2du^2+\beta^2dv^2$ and the activated (reference) metric becomes $ds^2_A=\lambda_1(\bL,\be)^2\alpha^2du^2+\lambda_2(\bL,\be)^2\beta^2dv^2$, where $\lambda_1$ and 
$\lambda_2$ can in general depend on $(u,v)$ through the designable material features $\be=\be(u,v)$.
If the activated sheet is blocked from deforming, we have
\begin{equation}
    g = \begin{pmatrix}
    \alpha(u,v)^2 & 0 \\
    0 & \beta(u,v)^2
    \end{pmatrix} ,
    \quad
    \bar{g} = \begin{pmatrix}
    \lambda_1(u,v)^2\alpha(u,v)^2 & 0 \\
    0 & \lambda_2(u,v)^2\beta(u,v)^2
    \end{pmatrix}. 
\end{equation}
Substituting into Eq.~\eqref{trace SM}, we obtain
\begin{equation} \label{energy 1}
E= h \underset{\text{Target Surface}}{\iint} \mathcal{E}(\lambda_1,\lambda_2) dA,
\end{equation}
where
\begin{equation} \label{density 1}
\mathcal{E}(\lambda_1,\lambda_2)= \frac{Y}{8(1-\nu_e^2)}\left[(1-\nu_e)\left(\left(1-\frac{1}{\lambda_1^2}\right)^2+\left(1-\frac{1}{\lambda_2^2}\right)^2\right)+\nu_e\left(\frac{1}{\lambda_1^2}+\frac{1}{\lambda_2^2}-2\right)^2\right],
\end{equation}
and $dA=\lambda_1\lambda_2\alpha\beta du dv$ is the area measure on the target surface.

As an illustrative example, we consider the case of a baromorph with $\lambda_1=1$ and $\lambda_2=1+\Lambda \epsilon$, where $\Lambda$ denotes pressure and $\epsilon$ is a designable feature related to the channel thickness \cite{BRRS19}.
Substituting into Eq.~\eqref{energy 1}, we obtain
\begin{equation}
E=\frac{hY}{8(1-\nu_e^2)}\int \frac{\epsilon^2\Lambda^2(2+\Lambda \epsilon)^2}{(1+\Lambda \epsilon)^3} \alpha\beta dudv.
\end{equation}
If the baromorph is homogeneous so that $\epsilon$ is constant, the elastic energy takes the simplified form
\begin{equation} \label{energy 3}
E=\frac{hA_0Y}{8(1-\nu_e^2)}\frac{\epsilon^2\Lambda^2(2+\Lambda \epsilon)^2}{(1+\Lambda \epsilon)^3},
\end{equation}
where $A_0$ is the area of the unactuated sheet.

In the main text, we consider a system with two degrees of freedom in actuation whose metric is given by 
\begin{equation}
ds^2=(1+\Lambda_2 \epsilon)^2v^2du^2 + (1+\Lambda_1 \epsilon)^2dv^2,
\end{equation}
where $\epsilon$ is taken to be a constant. As depicted in Figure 4 (d) of the main text, a 3-stage cycle is executed by taking $(\Lambda_1,\Lambda_2) = (0,0) \rightarrow (1,0) \rightarrow (1,1) \rightarrow (0,0)$ along straight lines in parameter space. The elastic energy in the frustrated system in stage I is simply Eq.~\eqref{energy 3} with $\Lambda = \Lambda_1$. In stage II, the elastic energy can be calculated by taking the reference metric as $ds^2=(1+\Lambda_2 \epsilon)^2v^2du^2 + (1+\epsilon)^2dv^2$ and varying $\Lambda_2$ from 0 to 1. Finally, the elastic energy in stage III is found by taking the reference metric to be
$ds^2=(1+\Lambda \epsilon)^2v^2du^2 + (1+\Lambda \epsilon)^2dv^2$ where $\Lambda = \Lambda_1 = \Lambda_2$, which yields
\begin{equation}
E=\frac{hA_0Y}{4(1-\nu_e)}\frac{\epsilon^2\Lambda^2(2+\Lambda \epsilon)^2}{(1+\Lambda \epsilon)^2}.
\end{equation} 
These energy calculations are combined to produce the plot in Figure 4 (d) of the main text, which further demonstrates the non-reciprocal nature of the cycle.

Finally, we note that while the non-Euclidean plate theory used above is valid for large displacements but small strains, it can be extended to systems involving large strains using suitable hyperelastic models, such as Mooney-Rivlin or neo-Hookean models. For instance, the corresponding stretching energy for a thin membrane of incompressible neo-Hookean elastomer takes the form
\begin{equation}
E=h \underset{\text{Target Surface}}\iint\frac{Y}{4(1+\nu_e)}\left[\operatorname{Tr}\left(\bar{g}^{-1}g\right)+\frac{1}{\det\left(\bar{g}^{-1}g\right)}-3\right]\frac{dA}{\lambda_1\lambda_2}
\end{equation}
and can be used to describe elastomers experiencing large strain deformations \cite{Duffy2020}. Repeating the earlier analysis with this stretch energy, $\mathcal{E}$ is replaced by
\begin{equation} \label{density2}
\mathcal{E}(\lambda_1,\lambda_2)= \frac{Y}{4(1+\nu_e)}\left(\frac{1}{\lambda_1^2}+\frac{1}{\lambda_2^2}+\lambda_1^2\lambda_2^2-3\right)\frac{1}{\lambda_1\lambda_2},
\end{equation}
which agrees with Eq.~\eqref{density 1} for small strains to quadratic order.


%

\end{document}